\documentclass[aps,prb,preprint,superscriptaddress,longbibliography]{revtex4-1}
\usepackage{amssymb}
\usepackage[utf8]{inputenc}
\usepackage{graphicx}
\usepackage{bm,color,subfigure,amsmath,longtable}

\begin{document}

\title{Topologically non-trivial magnon bands in artificial square spin ices subject to Dzyaloshinskii-Moriya interaction}

\author{Ezio~Iacocca}
\email{ezio.iacocca@colorado.edu}
\affiliation{Department of Applied Mathematics, University of Colorado, Boulder, Colorado 80309, USA}
\affiliation{Department of Physics, Division for Theoretical Physics, Chalmers University of Technology, 412 96, Gothenburg, Sweden}

\author{Olle~Heinonen}
\affiliation{Materials Science Division, Argonne National Laboratory, Lemont, IL 60439, USA}
\affiliation{Northwestern-Argonne Institute for Science and Engineering, Evanston, IL 60208, USA}

\begin{abstract}
Systems that exhibit topologically protected edge states are interesting both from a fundamental point of view as well as for potential applications, the latter because of the absence of back-scattering and robustness to perturbations. It is desirable to be able to control and manipulate such edge states. Here, we show that artificial square ices can incorporate both features: an interfacial Dzyaloshinksii-Moriya gives rise to topologically non-trivial magnon bands, and the equilibrium state of the spin ice is reconfigurable with different configurations having different magnon dispersions and topology. The topology is found to develop as odd-symmetry bulk and edge magnon bands approach each other, so that constructive band inversion occurs in reciprocal space. Our results show that topologically protected bands are supported in square spin ices.
\end{abstract}
\maketitle

\section{Introduction}

Topological insulators~\cite{Moore2007,Hsieh2008,Hsieh2009,Xia2009,Hsieh2009Science,Hasan2010} are generally materials that are insulating in the bulk but have conducting dissipationless edge states~\cite{Roushan2009}. In two dimensions, topological insulators (TIs) include quantum Hall (QH) states~\cite{Thouless1982PRL,Niu1985PRB}. QH edges states are metallic and chiral in that electrons on one physical edge move only in one direction; this prohibits back-scattering and makes the states dissipationless. The existence of topologically protected edge states is guaranteed if the band structure of the system has a non-trivial topology. A non-trivial topology is characterized by a non-zero Chern number, which is related to the Berry's phase that Bloch states $|u_{n,\vec {k}}\rangle$, where $\vec{k}$ is a wavevector in the first Brillouin zone and $n$ a band index, acquire when transported around a closed loop in the Brillouin zone; the Chern number is the total flux of the Berry's phase in the Brillouin zone. Therefore, in order for there to be a non-trivial topology, the Berry's phase accumulated around a closed loop cannot be zero. This may happen, but is not assured, if time-reversal invariance is broken.  

Topologically protected edge states are of great interest for potential applications, for example in information technology or communication systems. This extends beyond electronic systems, and includes photonic TIs~\cite{Khanikaev2013,Rechtsman2013} as well as certain magnonic crystals~\cite{Shindou2013,Shindou2013b}. In these latter systems, the band structure of spin excitations, or magnons, exhibit a topological order with non-trivial Chern number and protected edge states determined by the materials set and structure of the systems. However, systems with potential practical applications should be reconfigurable, so that the band structure of excitations can be modified with some external control parameter.

Artificial spin ices (ASIs)~\cite{Wang2005,Nisoli2013,Heyderman2013} are systems that allow for such reconfiguration. ASIs are composed of geometrically placed magnetic nanoislands coupled through dipolar interactions. These interactions stabilize the nanoislands' magnetization in configurations such that the magnetization at the lattice vertices satisfy an ``ice rule'' in low-energy states. ASIs are geometrically frustrated by design, that is, not all interactions at a given vertex can be simultaneously minimized. This leads to complex energy surfaces with many local energy minima. Consequently, significant efforts have been devoted to control and manipulate the magnetization state either by thermal or applied field protocols~\cite{Farhan2013} or novel geometries~\cite{Gilbert2014,Gilbert2015,Wang2016,Ma2016,Perrin2016}. From a dynamic perspective, ASIs compose a super-lattice that can be considered a magnonic crystal and therefore exhibit a rich band structure~\cite{Demokritov2013,Nikitov2001,Neusser2009,Kruglyak2010,Lenk2011,Krawczyk2014,Chumak2015}. This makes ASIs a natural system in which to explore reconfigurable magnonics~\cite{Grundler2015}, where the properties of the spin wave band structure can be actively controlled~\cite{Gliga2013,Gliga2015,Iacocca2016,Jungfleisch2016,Bhat2016,Zhou2016,Wang2017} to achieve functionality, chiefly for miniaturized microwave electronics~\cite{Kithun2010,Karenowska2012,Obry2013,Klinger2014,Vogel2015}.

A promising geometry for reconfigurable magnonics are square ASIs, where the magnetic nanoislands are placed at the sites of a square lattice. As a super-lattice, these are similar to arrays of dipolarly-coupled nanodots~\cite{Tacchi2011,Krawczyk2014} for which analytical methods calculating band diagrams under a macrospin approximation have been developed~\cite{Bondarenko2010,Verba2012,Lisenkov2016}. However, an initial micromagnetic study by Gliga et al.~\cite{Gliga2013} suggested that frustration in square ASIs modifies magnon modes by the existence of underlying defects known as Dirac strings~\cite{Heyderman2013}. These defects originate as the ice rule is locally broken, yet conserving the overall topological charge of the system. More recently, experiments in extended square ASIs lattices~\cite{Jungfleisch2016} demonstrated that the magnetization ground state determines the features and eigenfrequencies of the magnon modes. This conclusion was supported by a tight-binding-inspired semi-analytical model~\cite{Iacocca2016} that captures the dominant dipole-dipole, long-range coupling between the nanoislands, and thus provides a means to numerically compute the square ASI's band structure, otherwise impractical by more accurate models, e.g., micromagnetic simulations.

Square ASIs, though reconfigurable magnonic crystals, have magnon band structures with trivial topologies. It is of great interest to devise magnonic crystals that have non-trivial topological order \emph{and} that are reconfigurable. One way to realize such systems is to introduce interfacial Dzyaloshinskii-Moriya interactions (DMI)~\cite{Dzyaloshinskii1958,Moriya1960} to ASIs. DMI generally manifests as a chiral magnetic interaction in three-dimensional systems with broken inversion symmetry and can give rise to topological edge states in pyrochlores~\cite{Zhang2013} and spin textures in uniaxial thin film ferromagnets such as skyrmions~\cite{Rossler2006,Jiang2015,Heinonen2016}. More conveniently, an interfacial DMI can arise when a trivial magnet is deposited as a thin film on a strong spin-orbit scatterer, such as Ta or Pt. This effect has been used experimentally~\cite{Jiang2015} and numerically~\cite{Sampaio2013,Zhou2015,Heinonen2016} to nucleate and dynamically drive skyrmions at room temperature. Interfacial DMI naturally has a thickness-dependent strength~\cite{Nembach2015}, parametrized by an interfacial energy $D$ in units of J/m$^2$. Although most of the research on DMI has focused on topological structures or non-reciprocal spin wave dispersion in extended films~\cite{GarciaSanchez2014,Nembach2015}, the effect of DMI for spin waves in magnetic nanoislands has been studied only recently~\cite{Zingsem2016}. In the context of magnonic square ASIs, the addition of a chiral interfacial DMI suggests the possibility of topological magnon modes and novel features in their band structure.

The purpose of our work is to demonstrate that square ASIs subject to interfacial DMI admit topologically non-trivial bands, analogous to electronic TIs, and topologically trivial bands toggled only by the underlying magnetic configuration. In a magnonic system without DMI, the magnons are elliptical revolutions of the magnetization about its local equilibrium direction, and modes at $(\pm\vec{k},n)$ are degenerate. The form of the interfacial DMI breaks this degeneracy\cite{Zingsem2016} as it gives rise to an effective magnetic field 
\begin{equation}
\label{eq:Hdmi}
  \vec{H}_\mathrm{DMI} = \frac{2D}{M_s}\left[\left(\nabla\cdot\vec{m}\right)\hat{z}-\nabla m_z\right]
\end{equation}
that couples differently to states at $\vec{k}$ and $-\vec{k}$ and therefore gives rise to a coherent Berry phase accumulation. We will show that by reconfiguring the equilibrium state of the lattice the non-trivial topology can be turned off. Moreover, and external, in-plane magnetic field offers another degree of control to toggle topological bands and their propagation direction.

\section{Semi-analytical model}

To compute the band structure of a square ASI, it is necessary to calculate the long-range, dipole-dipole mediated magnon dispersion as a function of the reciprocal wavevector $\vec{k}$. For each wavevector, the dispersion relation is obtained from small-amplitude perturbations of the Larmor equation
\begin{equation}
\label{eq:LL}
  \frac{\partial\vec{m}}{\partial t} = -\gamma\mu_0\vec{m}\times\vec{H}_\mathrm{eff},
\end{equation}
where $\gamma$ is the gyromagnetic ratio, $\mu_0$ is the vacuum permeability, $\vec{m}$ is the magnetization vector normalized to the saturation magnetization $M_s$, and $\vec{H}_\mathrm{eff}$ is the effective field that includes diverse physical effects. In order to obtain a meaningful dispersion relation, a minimal model for the effective field must include an external field, exchange coupling, anisotropy, and DMI within a nanomagnet as well as dipolar interactions between nanomagnets. Given the cubic decay of the dipolar interactions, solving Eq.~\eqref{eq:LL} with such an effective field composes a daunting task requiring massive computational resources. While such a study can be performed~\cite{Semenova2013}, it is attractive to formulate a minimal model that captures the relevant physics of the system and minimizes the computation time. This is especially important to explore the existence of band-inversion, which requires a sufficiently resolved band structure. There are many possibilities to tackle this problem such as utilizing under-resolved micromagnetics to reduce the computational overhead (similar to the atomic structures explored in Refs.~[\onlinecite{Shindou2013}] and [\onlinecite{Shindou2013b}]), extrapolate from simpler systems, e.g., in Ref.~[\onlinecite{Dvornik2011}], or utilize periodic boundary conditions (PBCs) to estimate the band structure, e.g., in Ref.~\onlinecite{Tacchi2011}. However, these methods neglect important physical effects such as anisotropy and the long-range interactions across a periodic lattice. Instead, a tight-binding-inspired semi-analytical model including the relevant physics for square ASIs was recently shown to yield good agreement with both micromagnetic simulations~\cite{Iacocca2016} and experiments~\cite{Jungfleisch2016}.

The semi-analytical method is based on Eq.~\eqref{eq:LL} and relies on the conserved amplitude of the magnetization vector, $|\vec{m}|=1$, to represent small-amplitude perturbations from a homogeneous state as complex amplitudes, $a$. This is achieved by performing a Holstein-Primakoff transformation on the magnetization vector
\begin{equation}
\label{eq:HP}
  a = \frac{m_\xi+im_\eta}{\sqrt{2M_s(M_s+m_\zeta)}},
\end{equation}
where $\vec{m}=(m_\xi,m_\eta,m_\zeta)$ such that $m_\zeta$ is parallel to the local equilibrium magnetization direction and ($m_\xi,m_\eta$) represent orthogonal, small perturbations~\cite{Slavin2009}. Substituting Eq.~\ref{eq:HP} into Eq.~\ref{eq:LL}, a Hamiltonian system of equations is obtained as a function of $a$ and $a^*$ following the procedure outlined in Ref.~\onlinecite{Slavin2009}. The resulting Hamiltonian model can be generalized for $2\mathcal{N}$ interacting complex amplitudes and their complex conjugates, $\underline{a}$ and $\underline{a}^*$, so that $\partial_t\underline{a}=i\partial_{\underline{a}^*}\mathcal{H}(\underline{a},\underline{a}^*)$ and $\partial_t\underline{a}^*=-i\partial_{\underline{a}}\mathcal{H}(\underline{a},\underline{a}^*)$, where $\mathcal{H}(\underline{a},\underline{a}^*)$ is a Hamiltonian matrix~\cite{Iacocca2016}. These equations can be rewritten as an eigenvalue problem by means of Colpa's grand dynamical matrix~\cite{Colpa1978}
\begin{equation}
\label{eq:eigen}
    \omega\underline{\psi} = \mathcal{H}\underline{\psi}=\begin{pmatrix}
                            \mathcal{H}^{(1,2)} & \mathcal{H}^{(2,2)} \\
                            \mathcal{H}^{(1,1)} & \mathcal{H}^{(2,1)} \\
                          \end{pmatrix}\underline{\psi},
\end{equation}
from which we obtain the eigenvalues $\omega$ and eigenvectors $\underline{\psi}$ corresponding to each augmented vector of complex amplitudes $[\underline{a}^T,\underline{a}^\dagger]$. Because of the periodic structure of the ASI, we can label the eigenvectors $\underline{\psi}$ by a wavector $\vec{k}$ in the first Brillouin zone, and a band index $n$. The Hamiltonian matrix $\mathcal{H}$ is related to the effective field via $\mathcal{H}=-\gamma\delta W/(2M_s)$, where $\delta W=-\int \vec{H}_\mathrm{eff}(\vec{M}) \cdot d\vec{M}$ is the energy functional. In Ref.~\onlinecite{Iacocca2016}, the Hamiltonian matrices for an external in-plane field as well as anisotropy, dipole-dipole, and exchange fields were derived. To minimize finite-size errors from the long-range dipole-dipole Hamiltonian matrix, the lattice is allowed to grow until the relative error is no greater than $10^{-6}$. For all $\vec{k}$, this corresponds to a lattice of $100\times100$ unit cells. The exchange interactions within the nanomagnet are also critical to correctly describe edge modes~\cite{Carlotti2014} that are manifested in the magnon dispersion~\cite{Gliga2015}. Because we are interested in the low-energy sector of the square ASI dynamics, the magnetic nanoislands are divided in $3$ macrospins coupled by an effective exchange strength, which has been shown to return a faithful representation of the lowest energy bulk and edge modes~\cite{Iacocca2016,Jungfleisch2016}.

Here, we extend the model to include DMI, which means we have to include the effective interfacial DMI field~\cite{Heinonen2016}, given by Eq.~(\ref{eq:Hdmi}). This field favors a chiral tilt of the perpendicular magnetization component that stabilizes helical order in extended films~\cite{Uchida2006}. Therefore, the model must consider a small-amplitude precession about an arbitrary direction of the unit sphere. The resulting Hamiltonian matrices for the effective interfacial DMI field are
\begin{subequations}
\label{eq:Hd}
\begin{eqnarray}
\label{eq:Hd11}
  \mathcal{H}_{DMI}^{(1,1)} &=&   D'\begin{pmatrix}
	                                V_1 & \mathcal{O} & \mathcal{O} & \mathcal{O} \\
									                \mathcal{O} & H_1 & \mathcal{O} & \mathcal{O} \\
									                \mathcal{O} & \mathcal{O} & V_1 & \mathcal{O} \\
									                \mathcal{O} & \mathcal{O} & \mathcal{O} & H_1 \\
	                            \end{pmatrix}\\
\label{eq:Hd21}
  \mathcal{H}_{DMI}^{(1,2)} &=&   D'\begin{pmatrix}
	                                V_2 & \mathcal{O} & \mathcal{O} & \mathcal{O} \\
									                \mathcal{O} & H_2 & \mathcal{O} & \mathcal{O} \\
									                \mathcal{O} & \mathcal{O} & V_2 & \mathcal{O} \\
									                \mathcal{O} & \mathcal{O} & \mathcal{O} & H_2 \\
	                            \end{pmatrix},
\end{eqnarray}
\end{subequations}
where $V_1^{i,j}$, $V_2^{i,j}$, $H_1^{i,j}$, and $H_2^{i,j}$ are $3\times3$ complex matrices relating the intra-island mangetization components, given in the appendix, and $\mathcal{O}$ are $3\times3$ zero matrices. The effective DMI parameter in our discrete, macrospin representation is given by
\begin{equation}
\label{eq:DMIeff}
  D' = \frac{\gamma D}{2 M_s t},
\end{equation}
where the inverse dependence on thickness reflects the interfacial nature of this effect. However, we stress that the field Eq.~\eqref{eq:Hdmi} is considered to be homogeneous across the thickness; this is a good approximation for thin nanoislands in which the magnetization at any point is uniform through the thickness.

Because the semi-analytical method relies on a second-order perturbation of a well-defined magnetization state, the effect of DMI can only be included as a deviation from such a state i.e., low DMI strengths. The initial equilibrium states are determined by energy minimization using Eq.~(\ref{eq:LL}) with an added Gilbert damping term, which we calculate from full-scale micromagnetic simulations as detailed below.

\section{Chern number}

The introduction of DMI in a square ASI breaks time-reversal invariance which suggests that topological modes may exist. To test for topology, we calculate the Chern number $C_n$ for band $n$, defined as
\begin{equation}
\label{eq:chern}
  C_n = \frac{1}{2\pi i}\int{\left[\partial_xA_y(\vec{k})-\partial_yA_x(\vec{k})\right]\mathrm{d}^2k}
\end{equation}
where $A_\mu=\langle\underline{\psi}(k),\partial_\mu\underline{\psi}(k)\rangle$ is the Berry connection, $\mu=x,y$, and the eigenmodes 
$\underline{\psi}$ belong to band $n$. A non-zero Chern number indicates that the band experiences inversion which, in a finite lattice, leads to topological edge modes. We stress that the total Chern number of the band structure, obtained by summing $C_n$ over the bands $n$, is conserved to zero. Therefore, any non-zero Chern number must be balanced with an opposite-signed Chern number.

The numerical computation of the Chern number is performed following the method given in Ref.~[\onlinecite{Fukui2005}]. This method relies on lattice gauge theory to calculate the Chern number in a discretized Brillouin zone, minimizing numerical artifacts that might lead to a non-integer Chern number.

\section{Equilibrium magnetization states: micromagnetic simulations}

\begin{figure}
  \centering \includegraphics[width=2.5in]{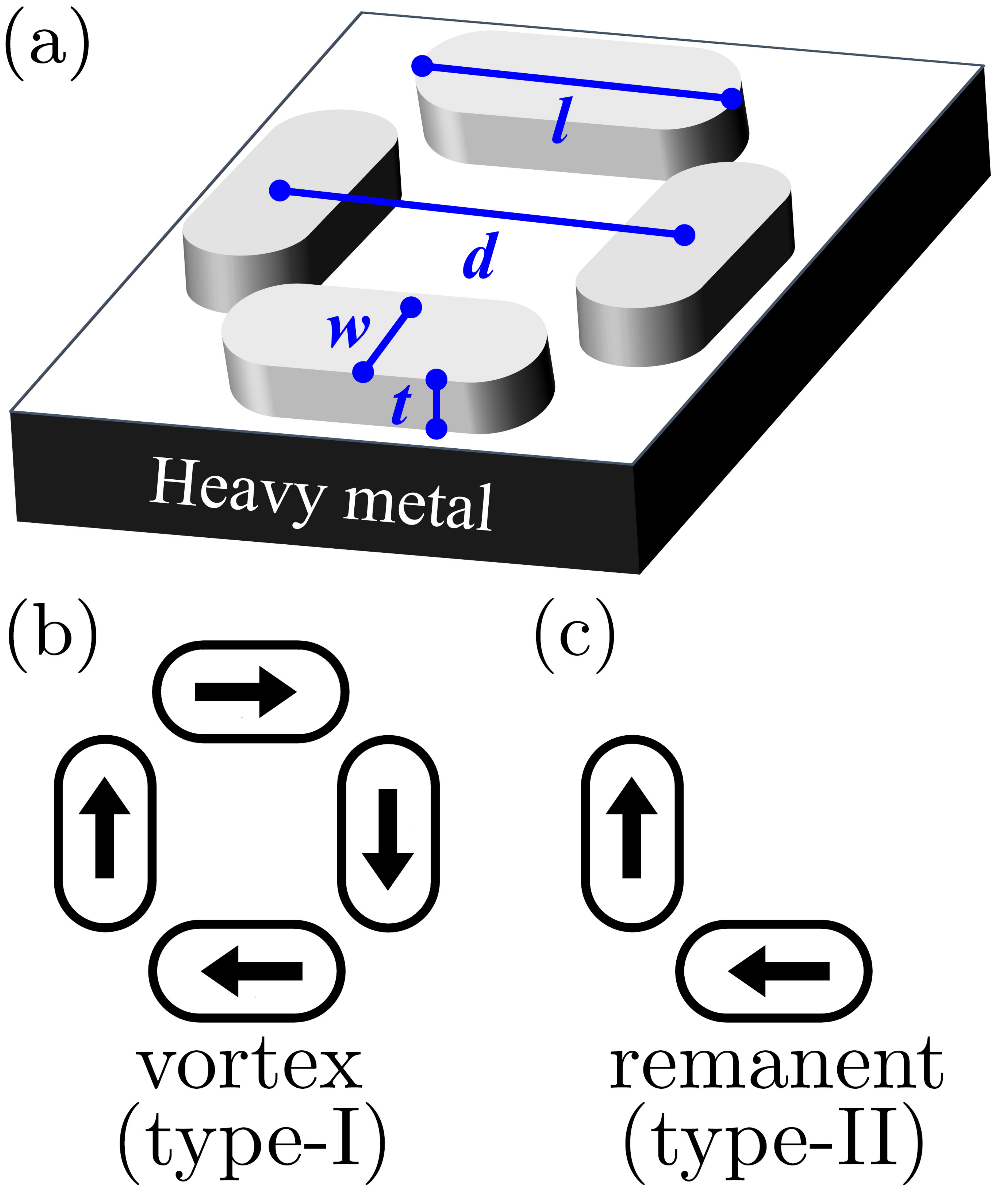}
  \caption{ \label{fig1} (a) Schematic of the square ASI unit cell. The composing nanoislands are identical with length $l$, width $w$, and thickness $t$. The lattice constant $d$ is taken as the center-to-center distance between adjacent nanoislands. Interfacial DMI is imparted through a heavy metal layer. (b) Schematic of a vortex (type-I) unit cell. (c) Schematic of a remanent (type-II) unit cell. }
\end{figure}

The semi-analytical model relies on determining the dispersion of small-amplitude modes about an equilibrium state. To determine these states in square ASIs subject to interfacial DMI, we perform micromagnetic simulations with both Mumax3~\cite{Vansteenkiste2014} and our in-house, double precision micromagnetic code used in e.g., Refs.~[\onlinecite{Heinonen2007,Schreiber2009,Muduli2011,Dumas2013,Heinonen2016}]. The square ASI unit cell and geometry is schematically shown in Fig.~\ref{fig1}(a), with lattice constant $d=390$~nm. The nanoisland are considered to be identical stadia, with lateral dimensions $l=290$~nm and $w=130$~nm, and variable thickness. These sizes are large enough for the nanoislands to support multiple modes. We consider two materials, Permalloy (Py) with $M_s=790$~A/m and Co$_{75}$Fe$_{25}$ with $M_s=1200$~kA/m, both with exchange stiffness $A=13$~pJ/m for simplicity. A heavy metal layer below the square ASI endows the  ferromagnetic nanoislands with an interfacial DMI. Numerically, we solve for a square ASI unit cell and impose periodic boundary conditions to simulate an extended lattice. To correctly account for the nanoislands' rounded edges in the micromagnetic finite difference scheme, we use a rather small cell size of $0.7$~nm in-plane and $5$~nm along the thickness~\footnote{We performed simulations with different in-plane cell size to show that for cells less than about $1.25$~nm in-plane, the simulations have converged and do not exhibit errors because of the ``staircase'' approximation of the rounded edges. We also performed simulations with a fixed in-plane cell size while varying the perpendicular dimension and found identical results for $5$~nm. The cell size of $0.7$~nm was chosen on the basis of numerical performance, so that the mesh has a number of cells proportional to a power of two}. Both micromagnetic codes returned identical ground states.

\begin{figure}
\centering \includegraphics[width=3.3in]{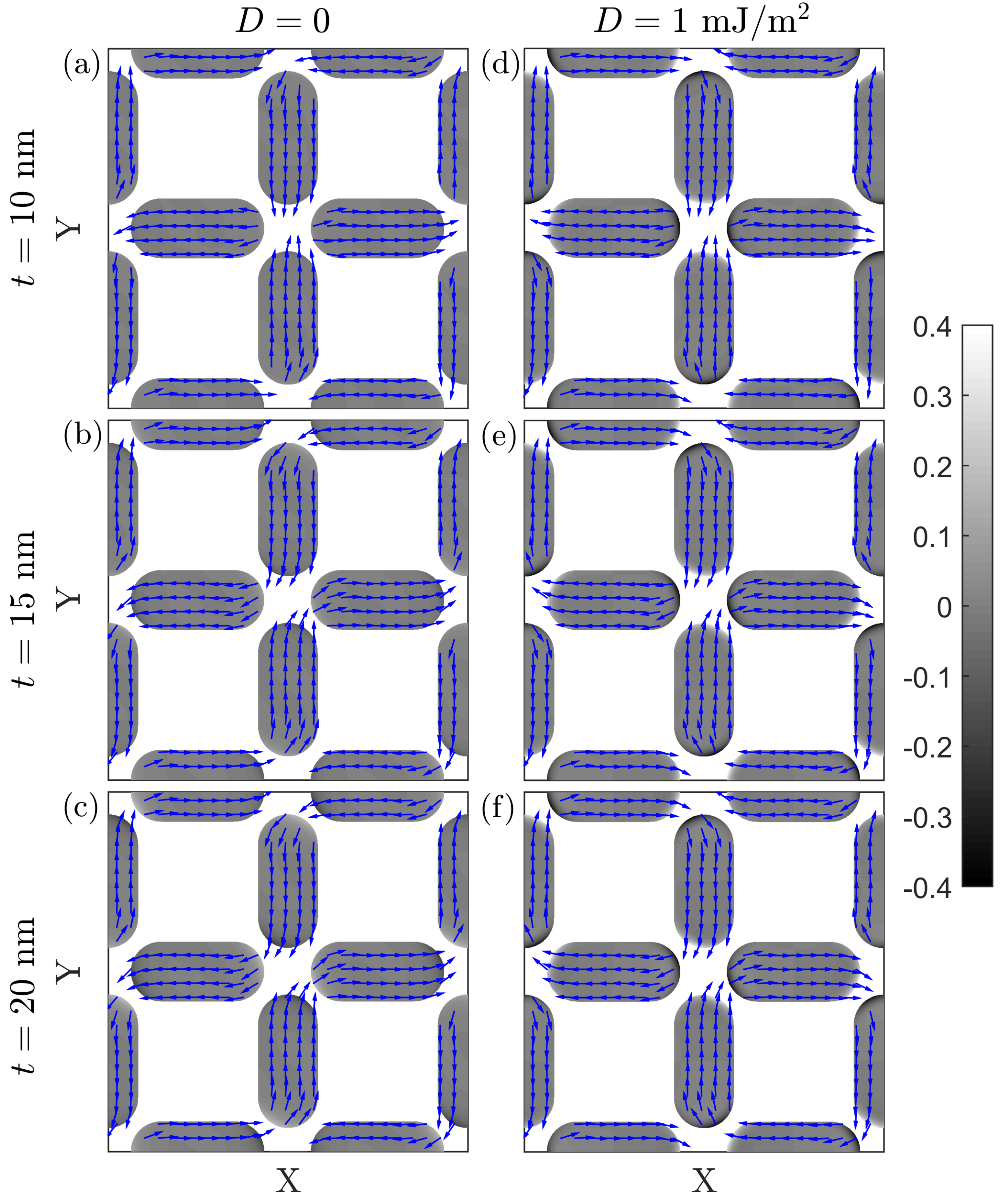}
  \caption{ \label{fig2} Micromagnetically computed ground states for a series of Py nanoislands of different thicknesses with (a-c) $D=0$ exhibiting S-states and (d-f) $D=1$~mJ/m$^2$ exhibiting C-states. The gray scale and arrows represent the $\hat{z}$ and in-plane magnetization components, respectively. }
\end{figure}

Two equilibrium configurations are computed: vortex (type-I) and remanent (type-II) states, schematically depicted in Fig.~\ref{fig1}(b) and (c), respectively. The simulation is initialized by setting a homogeneous magnetization in each nanoisland composing either of these states and allowing the simulation to relax using an artificially high Gilbert damping constant $\alpha=1$. In the remanent state, the equilibrium configurations are S-states and they are insensitive to DMI strength. In contrast, the vortex state exhibits a richer behavior as a function of $D$. For $D=0$, S-states are obtained, as shown in Fig.~\ref{fig2}(a-c) for the Py nanoislands of thicknesses $10$, $15$, and $20$~nm, where grayscale and arrows represent the $\hat{z}$ and in-plane magnetization component of the topmost layer, respectively. The inclusion of DMI favors C-states for all thicknesses, as shown for $D=1$~mJ/m$^2$ in Fig.~\ref{fig2}(d-f). Additionally, the DMI contributes to an out-of-plane tilt of the magnetization at the nanoislands' edges that is odd along the length of the island. However, this tilt is small and does not contribute significantly to the band structure. These simulations confirm that a moderate DMI only slightly perturbs the stable magnetization state in a square ASI, making it possible to study the full band structure by the semi-analytical model. We note that labyrinthine, chiral magnetization states are obtained for all thicknesses when $D\geq3$~mJ/m$^2$. For the Co$_{75}$Fe$_{25}$ nanoislands (not shown), a transition between a homogeneous, ``onion'' state and a C-state is observed as a function of $D$.

The spectrum at the $\Gamma$ point can be obtained micromagnetically by analyzing the equilibration after a weak perturbation of the system. We use a spatially homogeneous square field pulse as perturbation, with a duration of $50$~ps and a field magnitude of $10$~mT applied along the $(1,1)$ direction. The system is then relaxed for $10$~ns using a Gilbert damping of $\alpha=0.01$. This method only couples the uniform field to even modes but allows us to discern the dominant modes in one run. In the remanent state, constant eigenfrequencies are obtained as a function of $D$, as expected form the negligible impact of DMI on the equilibrium configuration. These eigenfrequencies are in agreement with those observed in Ref.~\onlinecite{Iacocca2016} and are topologically trivial, which we confirmed by calculating the Chern numbers of the magnon bands. In the vortex state, the $D$-dependent eigenfrequencies are shown in Fig.~\ref{fig3} for selected thicknesses of (a) Py and (b) Co$_{75}$Fe$_{25}$ nanoislands. The empty and filled red circles represent even bulk and edge modes, respectively. We note as a trend that the frequency drops as a function of $D$, consistent with the lower frequencies in the band diagram obtained from a C-state relative to an S-state~\cite{Iacocca2016}. It is also important to recognize that whereas we observe a single, fundamental bulk mode, the edge mode frequencies are split, consistent with the spin wave non-reciprocity induced by DMI and evidenced by a shift in their dispersion relation~\cite{Nembach2015,Zingsem2016}. However, we note that the frequency splitting is small, on the order of our numerical resolution of $24.5$~MHz.

\section{Band structure: semi-analytical calculations}

\begin{figure}
  \centering \includegraphics[width=3.3in]{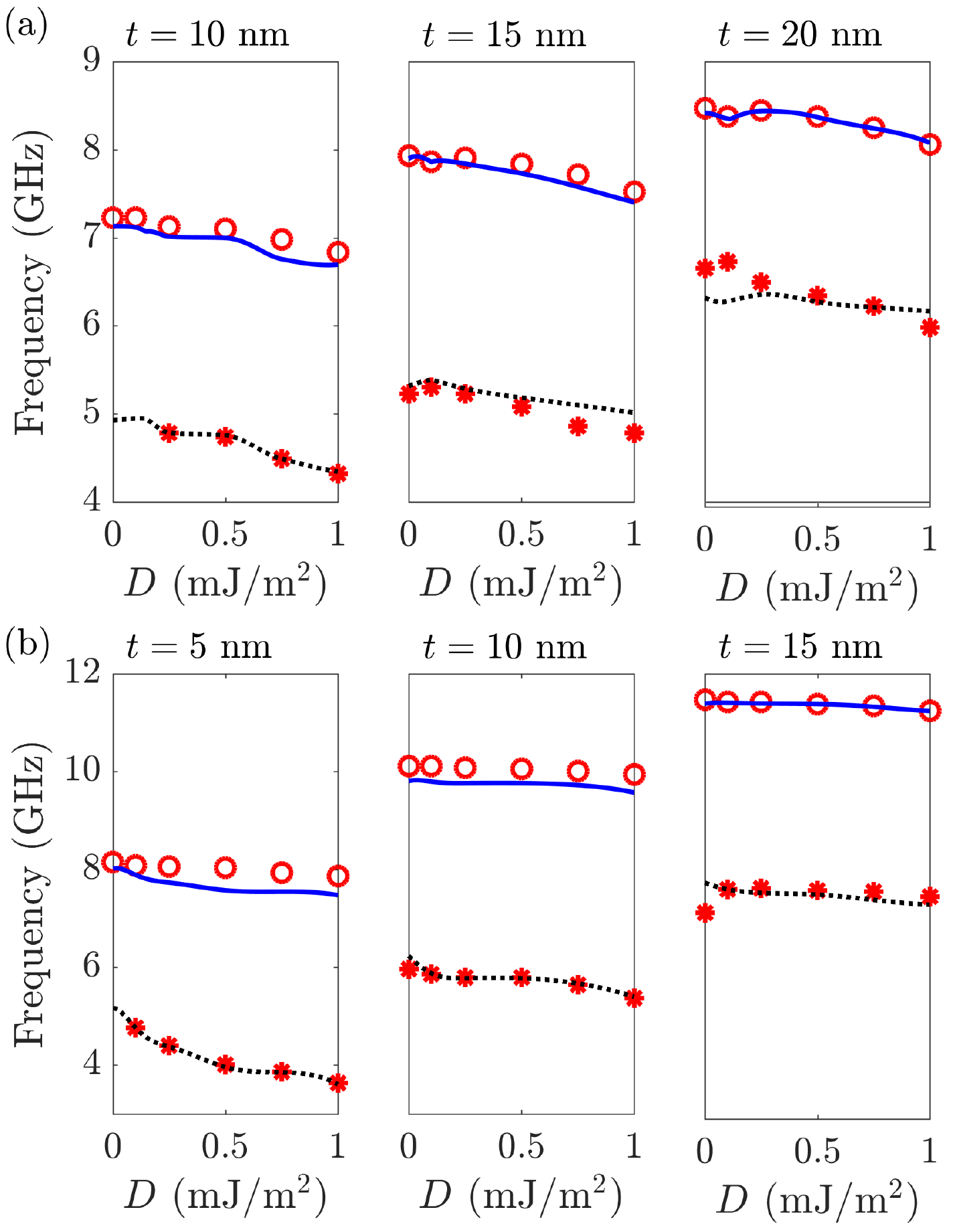}
  \caption{ \label{fig3} Even frequencies at the $\Gamma$ point as a function of $D$ for (a) Py nanoislands of thicknesses $t=10$, $15$, and $20$~nm and (b) Co$_{75}$Fe$_{25}$ nanoislands of thicknesses $t=5$, $10$, and $15$~nm. Good agreement is observed between the even bulk mode obtained semi-analytically (solid blue curves) and micromagnetically (empty red circles). The even edge mode obtained semi-analytically (black dashed curves) also agrees with those obtained micromagnetically (filled red circles). However, a qualitative deviation of the D-dependence for $20$~nm thick Py nanoislads (a, right panel) is observed. }
\end{figure}
Motivated by the micromagnetic simulations, we now utilize the semi-analytical model to solve for the eigenvalues in a square ASI with variable thickness and DMI. As a first step, we validate the semi-analytical model by finding the eigenvalues at the $\Gamma$ point as a function of $D$ in the vortex state. Because of the stadium-shape of the nanoislands, we adjust the anisotropy factors, estimated to first order by an ellipsoid~\cite{Osborn1945}. This is achieved by setting the equilibrium magnetization estimated from micromagnetic simulations and fitting the bulk and edge mode. For a finite $D$, we fit the eigenvalues by adjusting the in-plane tilt of the edge magnetization vectors in the semi-analytical model and assuming that all nanoislands in the unit cell behave identically. See Appendix B for the fitted parameters. Finally, we extrapolate the in-plane tilt of the edge magnetization vectors by a spline through the fitted points at $D=0$, $0.1$, $0.25$, $0.5$, $0.75$, and $1$~mJ/m$^2$. The resulting $D$-dependent eigenfrequencies at the $\Gamma$ point are shown in Fig.~\ref{fig3} for (a) Py and (b) Co$_{75}$Fe$_{25}$, where the solid blue and black dashed curves represent the even bulk and edge modes, respectively. Good qualitative agreement between micromagnetic simulations and semi-analytical calculations is obtained, suggesting that the semi-analytical model captures the relevant physics required to describe the dipole-mediated band structure including interfacial DMI. We note that the magnetization tilt required to fit the eigenfrequencies were below $25$~degrees in all cases, in agreement with the equilibrium states shown in Fig.~\ref{fig2}. In the case of $20$~nm thick Permalloy nanoislands, the calculated edge mode eigenfrequencies significantly deviate from those obtained micromagnetically. This is a consequence of spatial variations across the thickness of the nanoislands that are not taken into account semi-analytically. For $D>1$~mJ/m$^2$, DMI strongly perturbs the equilibrium state and dynamics at length scales much smaller than those captured by the three macrospins considered in the semi-analytical model ensue.

\begin{figure}[t!]
\centering \includegraphics[width=3.3in]{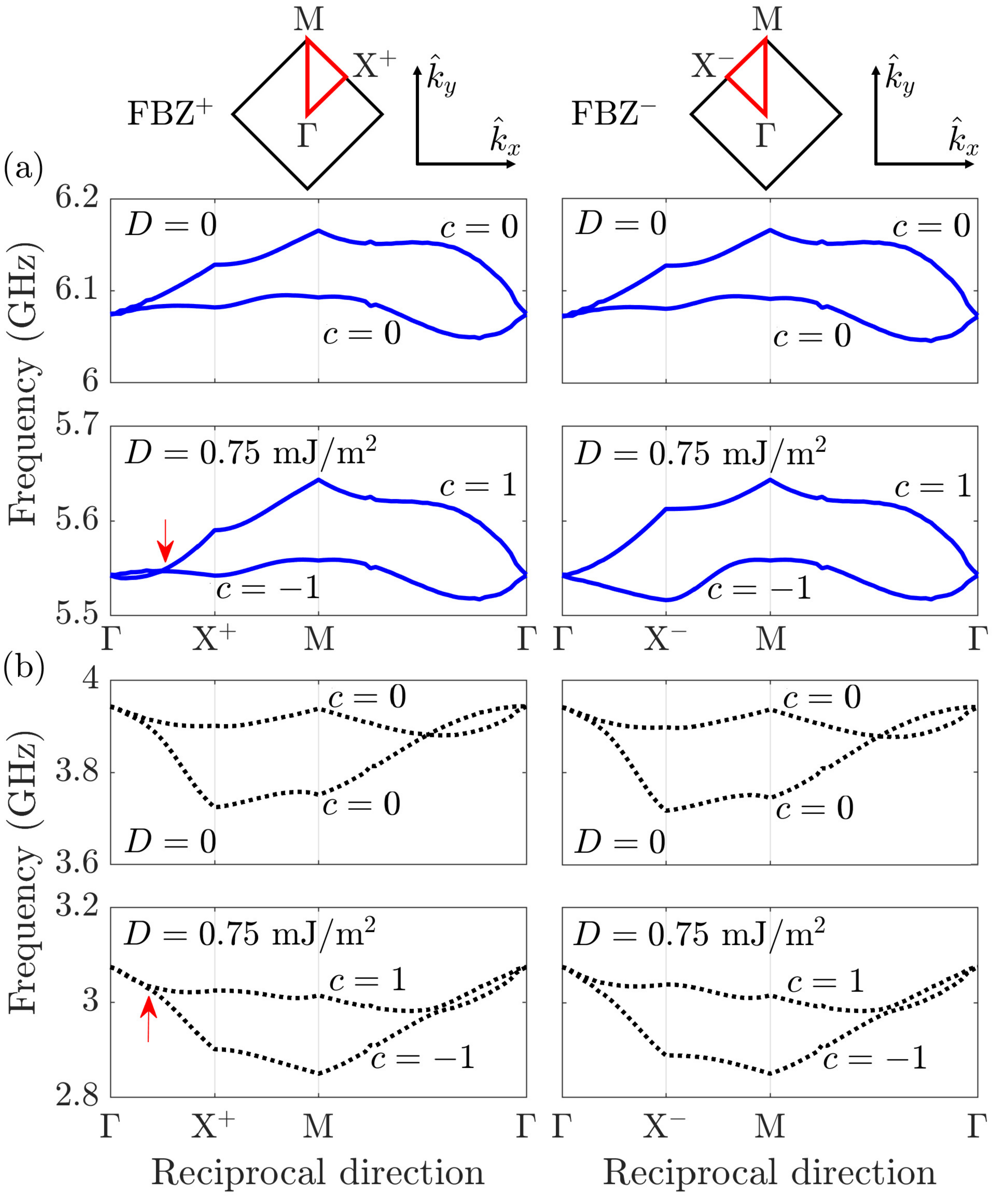}
  \caption{ \label{fig4} Band diagram for odd modes in $10$~nm-thick Py nanoislands. The FBZ path is depicted at the top of each column. Band diagrams for both $D=0$ and $D=0.5$~mJ/m$^2$ are shown. (a) Bulk odd modes exhibit a Dirac cone along the $\Gamma$-X direction, indicated by a red arrow. The symmetry between the paths FBZ$^+$ and FBZ$^-$ is broken when $D\not=0$; (b) Edge modes exhibit band touching in the $\Gamma$-M direction but for the edge mode bands, these do not accumulate Berry phase and the band touching occurs both for the topologically trivial and non-trivial cases.}
\end{figure}
It is worth noting that the semi-analytical model returns a total of twelve bands with three even- and nine odd-symmetry modes. These correspond to the four nanomagnets in the unit cell discretized in three exchange-coupled macrospins. As mentioned before, odd-symmetry modes cannot be excited micromagnetically with a homogeneous field; therefore, the semi-analytically-obtained odd modes are not shown in Fig.~\ref{fig3}.

We now calculate the band structure in the vortex state. For this, we compute the dispersion at $\vec{k}=(k_x,k_y)$ with $k_x$ and $k_y$ discretized in $0.05\pi/d$, composing a surface for each band $n$ in reciprocal space. The band-wise Chern number $C_n$ can be calculated by the method outlined in Sec. III. When $D=0$ we find $c=0$ for all thicknesses as expected from Ref.~\onlinecite{Iacocca2016}. However, for $D\neq0$, we find nonzero Chern numbers for both the bulk and edge odd-symmetry modes. This implies that DMI breaks the band structure inversion symmetry in reciprocal space.

As an example, we discuss below the first Brillouin zone (FBZ) band diagrams for $D=0$ (topologically trivial) and $D=0.75$~mJ/m$^2$ (topologically  non-trivial). The symmetry-breaking discussed above is most clearly seen for the bulk magnon bands by plotting the bands in FBZ$^\pm$ defined as a path through the $\Gamma$-X$^\pm$-M-$\Gamma$ directions, where the signs represent the relative sign of the $k_x$ and $k_y$ wavevector components, such that $X^\pm=\pi/(2d)(\hat{k}_x\pm \hat{k}_y)$. For $10$~nm thick Py nanoislands, we show in Fig.~\ref{fig4} the FBZ for the (a) bulk and (b) edge odd modes. For the bulk modes and with a DMI of $D=0.75$~mJ/m$^2$ and FBZ$^+$, the bands touch in (small) Dirac cones and exhibit inversion which, from a topological perspective, indicates constructive Berry phase accumulation and a non-zero Chern number. This is shown in Fig.~\ref{fig4}(a) along the $\Gamma$-X$^+$ direction, indicated by a red arrow. In contrast, the mirror path in FBZ$^-$ shown in Fig.~\ref{fig4}(a) right panels does not exhibit a Dirac cone for $D\not=0$, and the inequivalence of the two $X$-points in FBZ$^+$ and FBZ$^-$ is clearly exhibited. For the edge modes, the Dirac cones occur at a similar point along the $\Gamma$-X$^+$ direction for $D=0.75$~mJ/m$^2$, shown by the red arrow in Fig.~\ref{fig4}(b). Note that these bands do touch along the $\Gamma$-$M$ direction both for $D=0$ and $D\not=0$ but this does not contribute to the accumulated Berry's phase for $D\not=0$. Note that, as expected, the topologically non-trivial bands have oppositely signed Chern numbers, maintaining a trivial overall topology, so that the sum of the Chern numbers is zero. The same qualitative features are observed in the FBZ for $10$~nm thick Co$_{75}$Fe$_{25}$ nanoislads.

Topological bands appear for DMI strengths as low as $D=0.1$~mJ/m$^2$ in our simulations. This suggests that topology is a robust feature of the band structure, regardless of the discretization effects of our computation. However, we emphasize that the odd-symmetry bands lie within a range of $200$~MHz, representing a challenge for possible measurements at room temperature for these materials because of both intrinsic and extrinsic linewidth broadening that arise from damping and defects (e.g., edge inhomogeneities from patterning), respectively. It may be possible to resolve the bands using, e.g., meanderline resonance absorption~\cite{tsai2009}, provided the measured peaks have very good Lorentzian lineshape so that careful fitting will resolve them. As we discuss in the next section, this issue may be circumvented by applying an external field that both separates the bands and induces Dirac cones along specific directions.

In stark contrast to the qualitative features discussed above, the band diagram for the remanent state, see Fig.~\ref{fig1}(c), is topologically trivial as a function of $D$. This implies that it possible to toggle between topological and non-topological modes in a square ASIs by configuring the underlying magnetization configuration. In other words, ASIs can be utilized as a magnonic crystal with reconfigurable topological bands.

We stress that topologically non-trivial bands arise due to the broken degeneracy of $\vec{k}$ and $-\vec{k}$ states in the vortex configuration, mediated by interfacial DMI, that allows for a coherent Berry phase accumulation. The accompanying band inversion is a general feature of topologically non-trivial bands and can be further tuned by both material-specific parameters, e.g., saturation magnetization, and geometrical parameters, e.g., nanoislands shape and lattice constant $d$ and other types of ASIs~\cite{Gilbert2014,Gilbert2015,Perrin2016}.

\section{External field dependence}

An external magnetic field can tune the band frequency both by varying its magnitude, $|H|$, and angle, $\theta_H$, as previously shown for topologically trivial states~\cite{Iacocca2016}. Here, we explore the effect of an external, in-plane field on topologically non-trivial bands.

\begin{figure}
\centering \includegraphics[width=3.3in]{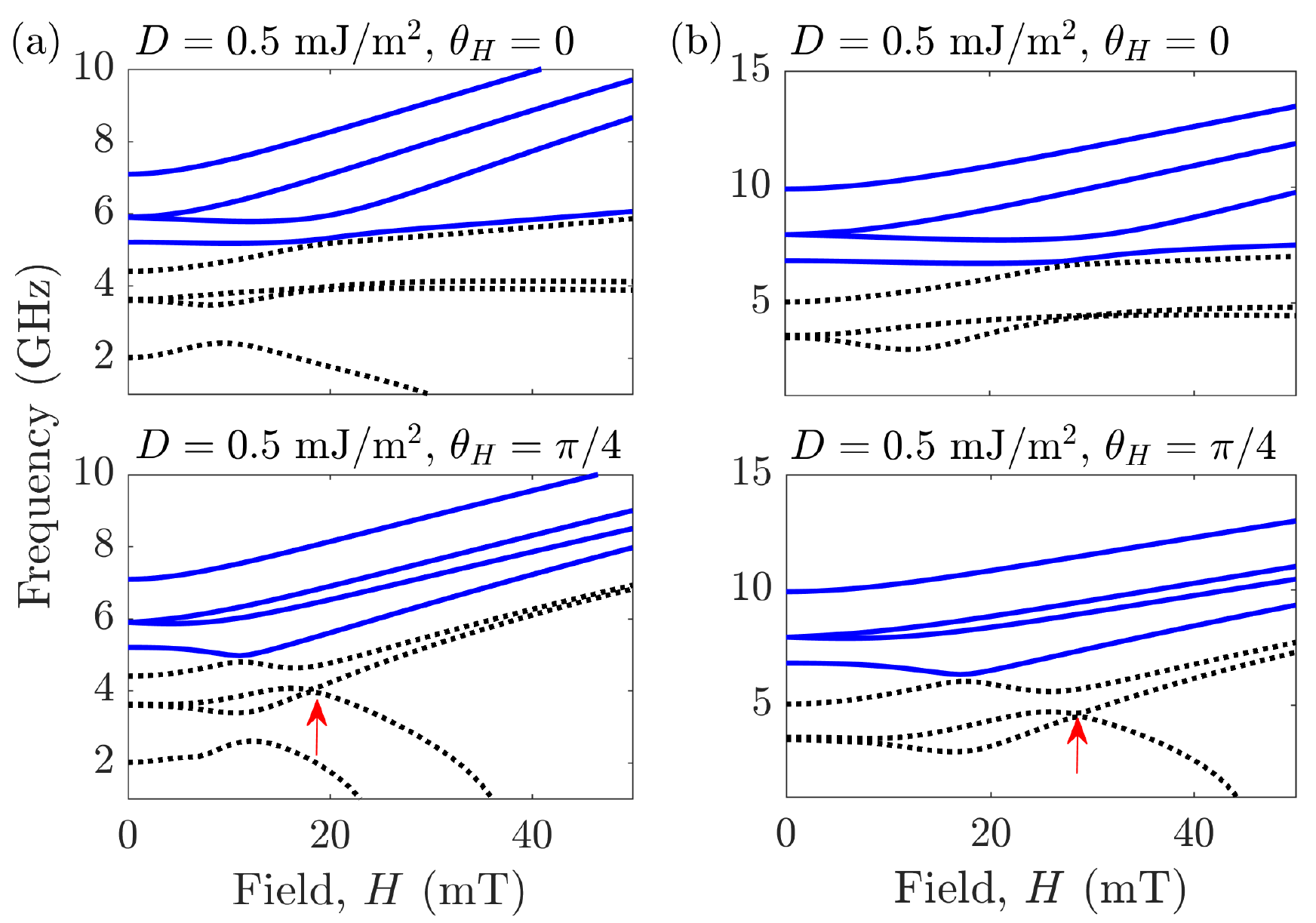}
  \caption{ \label{fig6} Field dependence of the band diagram at $\vec{k}=\pi/d(0.25, 0.25)$ for (a) Py and (b) Co$_{75}$Fe$_{25}$. The field is directed along the (1,0) direction in the top panel and along the (1,1) direction in the bottom panel. The red arrows indicate the appearance of a field-dependent Dirac cone at $18$~mT for Py and $28$~mT for Co$_{75}$Fe$_{25}$. }
\end{figure}

The broken symmetry induced by DMI suggests that the direction of the applied field can lead to significant changes in the band structure, including loss of topology. We explore the field magnitude dependence of the band structure at $\vec{k}=\pi/d(0.25, 0.25)$, the wavevector at which Dirac cones are observed for bulk modes in both Py and Co$_{75}$Fe$_{25}$ nanoislands of $10$~nm in thickness. For Py, the resulting field dependencies are shown in Fig.~\ref{fig6}(a) when the field is along the $(1,0)$ direction ($\theta_H=0$, top panel) or the $(1,1)$ direction ($\theta_H=\pi/4$, bottom panel). The bulk bands (blue curves) separate with field and mostly blue shift. However, the odd-symmetry edge exhibit a more complex behavior with field magnitude and angle. Notably, bands touch for a field of $18$~mT along the $\theta_H=\pi/4$ direction, indicated by a red arrow. Computing the full band structures at these conditions (not shown) indicates that the bands touch in a Dirac cone and the odd-symmetry edge modes become topologically protected. A similar field dependence is observed for Co$_{75}$Fe$_{25}$, shown in Fig.~\ref{fig6}(b). In this case, the field magnitude required to induce a Dirac cone in the odd-symmetry edge modes is $28$~mT, consistent with the higher saturation magnetization of Co$_{75}$Fe$_{25}$.

\begin{figure}
\centering \includegraphics[width=3.3in]{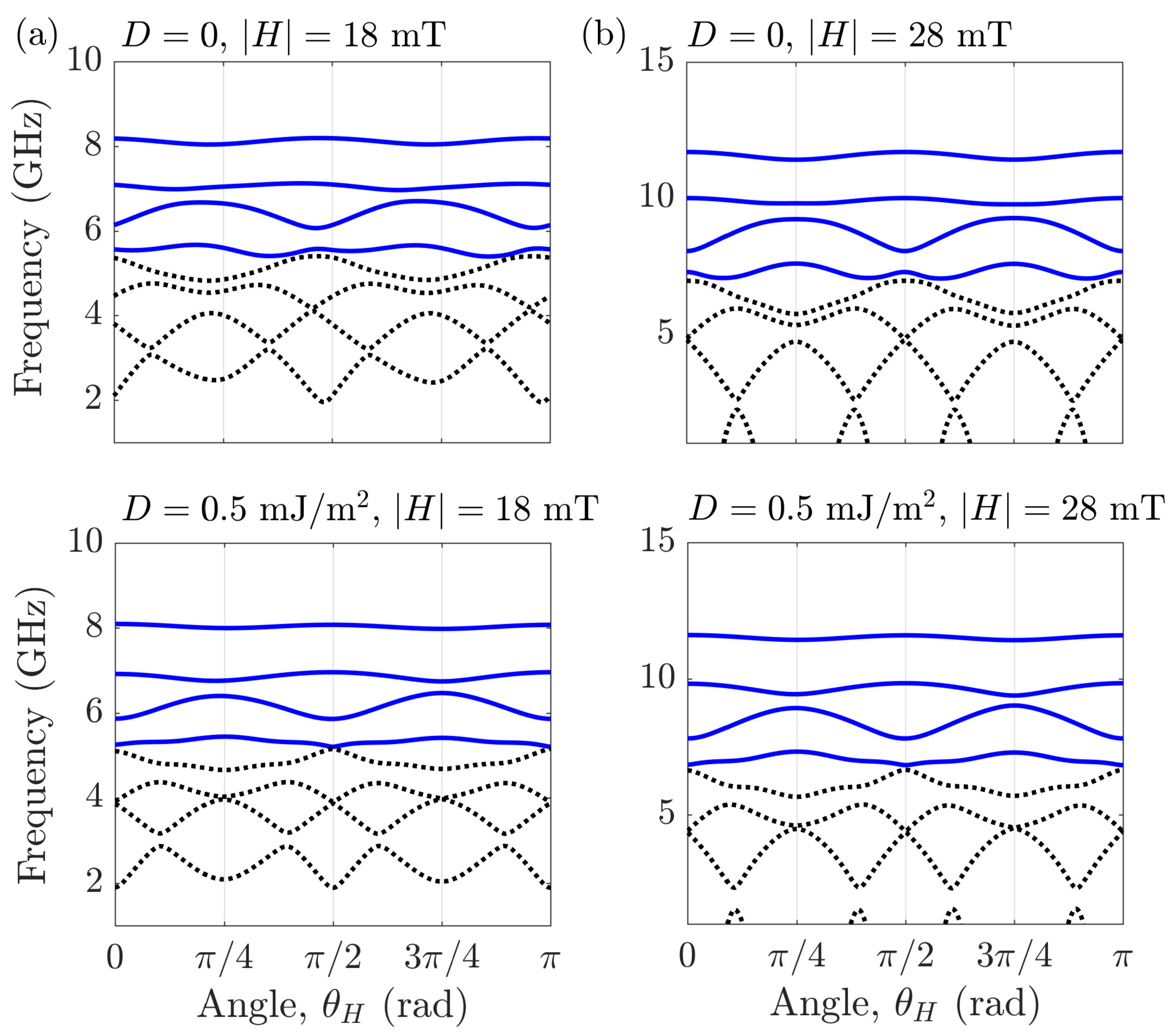}
  \caption{ \label{fig7} Angle dependence of the band diagram at $\vec{k}=\pi/d(0.25, 0.25)$ for (a) Py at $|H|=18$~mT; and (b) Co$_{75}$Fe$_{25}$ at $|H|=28$~mT. Bands do not touch when $D=0$ (top panels) while Dirac cones are observed in the odd-symmetry edge modes at $\theta_H=\pi/4$ and $3\pi/4$ (bottom panels). }
\end{figure}

To investigate the onset of Dirac cones at a finite field in more detail, we show in Fig.~\ref{fig7}(a) the band structure at $|H|=18$~mT for Py nanoislands and varying in-plane angle. In the absence of DMI (top panel), the bands do not touch at any angle. In particular, the bandgap between the odd edge modes is approximately $0.5$~GHz. In contrast, for $D=0.5$~mJ/m$^2$, we observe that Dirac cones appear at $\pi/4$ and $3\pi/4$, whereas the bands touching at $0$ and $\pi$ do not accumulate Berry phase. Because the edge modes now span a frequency range of approximately $1$~GHz as a function of angle, this method would allow one to experimentally measure Dirac cones in square ASIs. Similar qualitative results are observed for Co$_{75}$Fe$_{25}$, shown in Fig.~\ref{fig7}(b), where the bandgap is $1$~GHz with $D=0$ and frequency span of the odd edge modes is $5$~GHz.

\section{Conclusions}

In summary, we have calculated the spin wave band structure for square ASIs taking into account interfacial DMI imparted, e.g., by an adjacent heavy metal layer. The chiral nature of the DMI influences the eigenmodes supported by the square ASI, leading to band inversion through the development of Dirac cones. Our findings constitute a demonstration that magnon nonreciprocity within a magnetic nanoisland can be manifested at longer wavelengths through dipole coupling, leading to topologically protected edge modes in square ASIs. The magnon bands arise from long-range magnetostatic interactions between modes in individual islands. Only magnon bands that have odd spatial symmetry at the Brioullin zone center ($k=0$) develop topologically non-trivial modes, consistent with the fact that these can more efficiently couple with neighboring spins. It is also observed that topological bands establish a preferred propagation direction that corresponds to the non-reciprocity imparted by DMI.

The topologically protected magnon bands mentioned above suggest that square ASIs can withstand both thermal fluctuations and magnon scattering events. This is especially important for magnonic applications where spin waves are required to travel long distances in order to achieve logic and data transfer functionality within an all-magnetic circuitry. Furthermore, these features can be reconfigured and the non-trivial band topology -- and concomitantly topologically protected edge states -- turned off by changing the underlying magnetization configuration of the square ice, i.e., by different field protocols~\cite{Jungfleisch2016} to relax the nanoislands' magnetization. For instance, one can envision logic circuits based on the preferred propagation direction of topologically protected waves, toggled by the reconfiguration of a handful of nanoelements that would act as a tunable gate. It is also possible to envision modes propagating at the physical edges of the square ASI lattice exhibiting a much lower decay to magnetic damping based on the non-zero Chern number. However, a finite-sized, discrete lattice can strongly affect the dispersion of surface waves and a detailed study is required to assess the existence of true (physical) edge modes.

It is noteworthy that topology ensues as the bands approach each other in frequency, making it a challenging measurement due to the spectral broadening arising because of thermal fluctuations at finite temperatures, and spectral mixing. An alternative is to increase the band separation by utilizing an in-plane magnetic field and perform magnitude- and angle- dependent measurements to find evidence of Dirac cones at finite wavevectors. A plausible method to detect the resulting features at finite wavevectors is to use a meander line patterned on top of the square ASI as an antenna with $10$~MHz resolution and carefully deconvoluting spectral mixing to discern between the two broad spectral features. By measuring the bands in square ASIs as a function of the spin-orbit scatterer material and the thickness of the magnetic material, it would be possible to experimentally determine the onset of topologically non-trivial bands.

\begin{acknowledgments}
E.~I. acknowledges support from the Swedish Research Council, Reg. No. 637-2014-6863. The work by O.~H. was funded by the Department of Energy Office of Science, Materials Sciences and Engineering Division. We gratefully acknowledge the computing resources provided on Blues, a high-performance computing cluster operated by the Laboratory Computing Resource Center at Argonne National Laboratory.
\end{acknowledgments}

\appendix
\section{Matrix components of the DMI Hamiltonian}

The Hamiltonian matrices are written in terms of the complex amplitudes $a$, which are related to the normalized magnetization vector through their spherical components i.e., the polar and azimuthal angles, $\theta=\pi/2$ and $\varphi$, respectively (See Ref.~\onlinecite{Slavin2009} and Ref.~\onlinecite{Iacocca2016} for details). By performing the cross product $\vec{m}_i\times\vec{m}_j$, where $i$ and $j$ are two neighboring macrospins in a nanoisland, and keeping terms to second order in $a$, we obtain the $3\times3$ matrices
\begin{subequations}
\begin{eqnarray}
\label{eq:V1}
  V_1 &=& \begin{pmatrix}
	           0         & R_v^{1,2} & 0         \\
						 R_v^{2,1} & 0         & R_v^{2,3} \\
						 0         & R_v^{3,2} & 0         
					\end{pmatrix},\\
\label{eq:V2}
  V_2 &=& \begin{pmatrix}
	           S_v^{1,2}+S_v^{2,1} & C_v^{1,2}            & 0         \\
						 C_v^{2,1}           & \sum S_v   & C_v^{2,3} \\
						 0                   & C_v^{3,2}  & S_v^{2,3}+S_v^{3,2} 
					\end{pmatrix},\\
\label{eq:H1}
  H_1 &=& \begin{pmatrix}
	           0         & R_h^{1,2} & 0         \\
						 R_h^{2,1} & 0         & R_h^{2,3} \\
						 0         & R_h^{3,2} & 0         
					\end{pmatrix},\\
\label{eq:H2}
  H_2 &=& \begin{pmatrix}
	           S_h^{1,2}+S_h^{2,1} & C_h^{1,2}            & 0         \\
						 C_h^{2,1}           & \sum S_h   & C_h^{2,3} \\
						 0                   & C_h^{3,2}  & S_h^{2,3}+S_h^{3,2}
					\end{pmatrix},
\end{eqnarray}
\end{subequations}
where $\sum S_v = S_v^{1,2}+S_v^{2,1}+S_v^{2,3}+S_v^{3,2}$, $\sum S_h =S_h^{1,2}+S_h^{2,1}+S_h^{2,3}+S_h^{3,2}$, and
\begin{subequations}
\begin{eqnarray}
\label{eq:Rv}
  R_v^{i,j} &=& \sin\theta_i\sin\varphi{i}\cos\theta_j-\sin\theta_j\sin\varphi_j\cos\theta_i\nonumber\\
	          & & +i\left(|\cos\varphi_j|\cos\theta_i-|\cos\varphi_i|\cos\theta_j\right)\\
\label{eq:Sv}
  S_v^{i,j} &=& 2\left(\cos\theta_j\sin\varphi_j\sin\theta_i-\cos\theta_i\sin\varphi_i\sin\theta_j\right)\\
\label{eq:Cv}
  C_v^{i,j} &=& \sin\theta_j\sin\varphi_j\cos\theta_i-\sin\theta_i\sin\varphi_i\cos\theta_j\nonumber\\
	          & & +i\left(|\cos\varphi_i|\cos\theta_j+|\cos\varphi_j|\cos\theta_i\right)\\
\label{eq:Rh}
  R_h^{i,j} &=& \sin\theta_i\cos\varphi{i}\cos\theta_j-\sin\theta_j\cos\varphi_j\cos\theta_i\nonumber\\
	          & & +i\left(|\sin\varphi_j|\cos\theta_i-|\sin\varphi_i|\cos\theta_j\right)\\
\label{eq:Sh}
  S_h^{i,j} &=& 2\left(\cos\theta_j\cos\varphi_j\sin\theta_i-\cos\theta_i\cos\varphi_i\sin\theta_j\right)\\
\label{eq:Ch}
  C_h^{i,j} &=& \sin\theta_j\cos\varphi_j\cos\theta_i-\sin\theta_i\cos\varphi_i\cos\theta_j\nonumber\\
	          & & +i\left(|\sin\varphi_i|\cos\theta_j+|\sin\varphi_j|\cos\theta_i\right)\\
\end{eqnarray}
\end{subequations}

\section{Fitted parameters for the semi-analytical model}

To fit the micromagnetically to the semi-analytically calculated eigenfrequencies, we perform a two-step fitting for the anisotropy factors and the magnetization tilt angles. The fitted anisotropy factors for Py and Co$_{75}$Fe$_{25}$ for each thickness are listed in table~\ref{tab:1} and compared to the anisotropy factors from an ellipsoid~\cite{Osborn1945}. The nomenclature $N$, $L$, $M$ is used for the out-of-plane, easy, and hard axis anisotropy factors. 
\begin{table}
\caption{\label{tab:1} Fitted anisotropy factors used in the semi-analytical calculations and the first-order estimates from an ellipsoid. }
\begin{ruledtabular}
\begin{tabular}{ l l l l }
Thickness & Py & Co$_{75}$Fe$_{25}$ & Ellipsoid~\cite{Osborn1945}\\
\colrule
$5$~nm  &             & $N$ = 0.957 & $N$ = 0.9573 \\
        &             & $L$ = 0.007 & $L$ = 0.0139 \\
				&             & $M$ = 0.036 & $M$ = 0.0287 \\
\colrule
$10$~nm & $N$ = 0.925 & $N$ = 0.930 & $N$ = 0.9146 \\
        & $L$ = 0.020 & $L$ = 0.014 & $L$ = 0.0279 \\
				& $M$ = 0.065 & $M$ = 0.056 & $M$ = 0.0575 \\
\colrule
$15$~nm & $N$ = 0.890 & $N$ = 0.860 & $N$ = 0.8720 \\
        & $L$ = 0.020 & $L$ = 0.037 & $L$ = 0.0418 \\
				& $M$ = 0.090 & $M$ = 0.103 & $M$ = 0.0862 \\
\colrule
$20$~nm & $N$ = 0.780 &             & $N$ = 0.8293 \\
        & $L$ = 0.058 &             & $L$ = 0.0557 \\
				& $M$ = 0.162 &             & $M$ = 0.1150 \\
\end{tabular}
\end{ruledtabular}
\end{table}

The angles are fitted for each nanoislands to obtain quantitative agreement between the micromagnetic and semi-analytic even bulk and edge modes. The fitted angles are listed in table~\ref{tab:2} for Py and table~\ref{tab:3} for Co$_{75}$Fe$_{25}$. To account for both S and C states, the angles are fitted for both the north and south macrospins, relative to the direction of the magnetization in each nanoisland. The fitted angles in all cases are similar to those obtained by averaging the magnetization angles of the micromagnetic ground states. For example, in the case of Py nanoislands of $15$~nm, micromagnetic simulations return north and south tilt angles of $13$ and $-13$ deg at $D=0.5$~mJ/m$^2$ comparable to the fitted north and south tilt angles of $15$ and $-15$ deg.
\begin{table}
\caption{\label{tab:2} Fitted north (N) and south (S) tilt angles for Py. }
\begin{ruledtabular}
\begin{tabular}{ l l l l }
$D$             & $10$~nm     & $15$~nm     & $20$~nm \\
\colrule
$0$~mJ/m$^2$    & N = -10 deg   & N = -15 deg   & N = -15 deg\\
                & S = -10 deg   & S = -15 deg   & S = -15 deg\\
\colrule
$0.10$~mJ/m$^2$ & N = -5 deg  & N = 0 deg  & N = -1 deg\\
                & S = -10 deg & S = -10 deg  & S = 15 deg\\
\colrule
$0.25$~mJ/m$^2$ & N = 12 deg  & N = 10 deg  & N = 5 deg\\
                & S = -12 deg & S = -10 deg & S = -15 deg\\
\colrule
$0.50$~mJ/m$^2$ & N = 13 deg  & N = 15 deg  & N = 10 deg\\
                & S = -13 deg & S = -15 deg & S = -10 deg\\
\colrule
$0.75$~mJ/m$^2$ & N = 22 deg  & N = 20 deg  & N = 15 deg\\
                & S = -22 deg & S = -20 deg & S = -15 deg\\
\colrule
$1.00$~mJ/m$^2$ & N = 25 deg  & N = 25 deg  & N = 20 deg\\
                & S = -25 deg & S = -25 deg & S = -20 deg\\
\end{tabular}
\end{ruledtabular}
\end{table}

\begin{table}
\caption{\label{tab:3} Fitted north (N) and south (S) tilt angles for Co$_{75}$Fe$_{25}$. }
\begin{ruledtabular}
\begin{tabular}{ l l l l }
$D$             & $5$~nm      & $10$~nm     & $15$~nm \\
\colrule
$0$~mJ/m$^2$    & N = 0 deg   & N = 0 deg   & N = 0 deg\\
                & S = 0 deg   & S = 0 deg   & S = 0 deg\\
\colrule
$0.10$~mJ/m$^2$ & N = 13 deg  & N = 10 deg  & N = 4 deg\\
                & S = -13 deg & S = -8 deg  & S = -4 deg\\
\colrule
$0.25$~mJ/m$^2$ & N = 15 deg  & N = 10 deg  & N = 6 deg\\
                & S = -15 deg & S = -10 deg & S = -6 deg\\
\colrule
$0.50$~mJ/m$^2$ & N = 20 deg  & N = 10 deg  & N = 7 deg\\
                & S = -20 deg & S = -10 deg & S = -7 deg\\
\colrule
$0.75$~mJ/m$^2$ & N = 23 deg  & N = 10 deg  & N = 10 deg\\
                & S = -23 deg & S = -12 deg & S = -10 deg\\
\colrule
$1.00$~mJ/m$^2$ & N = 25 deg  & N = 17 deg  & N = 13 deg\\
                & S = -25 deg & S = -17 deg & S = -13 deg\\
\end{tabular}
\end{ruledtabular}
\end{table}

\end{document}